\documentclass[prb,aps, twocolumn]{revtex4-1}
\usepackage{graphicx}
\usepackage{color}
\usepackage[tbtags]{amsmath}
\usepackage{amssymb}
\usepackage{bm}

\newcommand{\be}{\begin{equation}}
\newcommand{\ee}{\end{equation}}
\newcommand{\HH}{{\cal H}}

\newcommand{\p}{\partial}
\newcommand{\s}{\sigma}
\newcommand{\la}{\langle}
\newcommand{\ra}{\rangle}
\newcommand{\lb}{\left[}
\newcommand{\rb}{\right]}
\newcommand{\lp}{\left(}
\newcommand{\rp}{\right)}

\renewcommand{\Im}{{\textrm{ Im}\,}}
\renewcommand{\vec}[1]{{\bm #1}}

\renewcommand{\phi}{\varphi}
\renewcommand{\epsilon}{\varepsilon}

\renewcommand{\dag}{\dagger}

\begin{document}

\title{Photoexcitation Cascade and Quantum-Relativistic Jets in Graphene}

\author{Cyprian Lewandowski, L. S. Levitov}

\affiliation{Department of Physics, Massachusetts Institute of Technology, Cambridge, MA 02139, USA}

\begin{abstract}
In Dirac materials linear band dispersion blocks momentum-conserving interband transitions, creating a bottleneck for electron-hole pair production and carrier multiplication in the photoexcitation cascade. Here we show that the decays are unblocked and the bottleneck is relieved by subtle many-body effects involving multiple off-shell e-h pairs. The decays result from a collective behavior due to emission of many soft pairs. We discuss
characteristic signatures of the off-shell pathways, in particular the sharp angular distribution of secondary carriers, resembling relativistic jets in high-energy physics. 
The jets can be directly probed using solid-state equivalent of particle detectors. Collinear scattering enhances carrier multiplication, allowing for emission of as many as ${\sim}10$ secondary carriers per single absorbed photon.  
\end{abstract}

\maketitle

\begin{figure}
\centering
\includegraphics[width=1.05\linewidth]{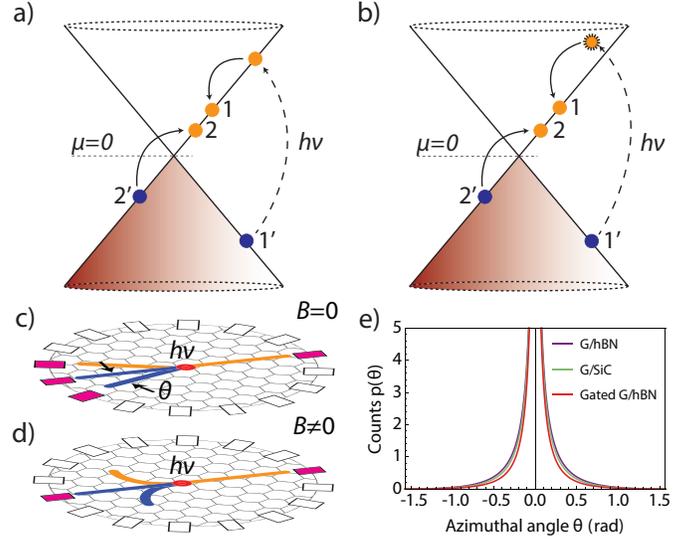}
\caption{
a,b) Types of carrier scattering in a Dirac band. The on-shell processes (a) are subject to energy and momentum conservation, and therefore cannot trigger transitions between physical states in different linearly dispersing bands \cite{PhysRevB.83.155441,PhysRevB.87.155429}. This bottleneck is relieved by the off-shell processes (b) mediated by virtual states residing off the Dirac cone.  This 
triggers collinear scattering and emission of multiple soft e-h pairs with a tightly focused jet-like angular distribution. The jets can be probed as illustrated in (c,d).
A photon (red dot) creates an e-h jet that is detected by a group of adjacent contacts (activated contacts are shown in magenta).
A weak 
$B$ field blocks soft pairs from reaching contacts (d), allowing for the energy distribution to be directly probed.
e) Angular distribution of 
soft pairs 
in the jets. The e-e interaction screened by the substrate and gate is described in [\onlinecite{Note1_bib}]. 
}
\vspace{-5mm}
\label{fig:jet_process}
\end{figure}

The general question of how an excited electron partitions its energy among lower-energy excitations is central to our understanding of carrier dynamics in solids.
One key pathway is the emission of particle-hole pairs, a process that leads to carrier multiplication in a photoexcitation cascade. Physics becomes particularly interesting in Dirac materials with linear carrier dispersion \cite{RevModPhys.81.109}, where strong interactions enhance the carrier-carrier scattering whereas momentum conservation greatly restricts the phase space available for such processes and (naively) may entirely block decays [see Fig.\ref{fig:jet_process}(a)]\cite{PhysRevLett.77.3589, PhysRevB.76.155431,PhysRevB.79.085415}. 

In models of 
photoresponse it is usually taken 
for granted that energy is conserved at all times and throughout all stages of the cascade, with transitions taking place `on-shell' \cite{PhysRevB.85.241404, 10.1038/ncomms2987, PhysRevB.88.035430, PhysRevB.87.155429, 0953-8984-27-16-164201, PhysRevB.94.205306}. Here we introduce the off-shell processes involving virtual states that disobey the energy-momentum relation. We argue that these processes dominate photoresponse, producing large numbers of secondary electron-hole (e-h) pairs. These processes are conceptually similar to the  off-shell processes in high-energy physics responsible for the formation of relativistic jets. 

The dilemma faced by a photoexcited electron in a Dirac material can be summarized through the quantum-mechanical uncertainty relation. The latter permits energy non-conservation for relatively short time intervals not exceeding the inverse decay time:
\be
\label{eq:offshellness_eq_1}
\Delta\epsilon\lesssim \frac{\hbar}{\tau}
.
\ee
Suppose the dependence $\tau$ vs. $\Delta\epsilon$ is such that increasing the ``offshellness" $\Delta\epsilon$ opens up a large phase space for decays. 
In this case, 
the off-shell processes with large $\Delta\epsilon$ will win 
over the processes with a smaller $\Delta\epsilon$.

As we will see, the offshell dynamics has striking consequences for the photoexcitation cascade and, ultimately, the photoresponse. First, it allows
primary photoexcited e-h pair to generate multiple secondary pairs, through the processes of the type pictured in Fig.\ref{fig:jet_process}(b). These pairs are typically considerably softer than the primary pair, forming a broadband energy distribution analyzed below.  Second, due to the collinear character of relevant electron-electron (e-e) collision processes, the secondary pairs are preferentially emitted along the primary pair velocity direction, forming a jet-like angular distribution [see Fig.\ref{fig:jet_process}(c)-(e)]. The latter can be studied experimentally using a solid-state analog of a particle detector realized as a circular array of photocurrent detectors \cite{nat_nano_2012_sun, nat_phys_graham, 0953-8984-27-16-164207}, see Fig.\ref{fig:jet_process}(c). 

Energy-resolved studies of soft pairs can be performed using an external magnetic field of strength such that it deflects the orbits of soft carriers but has little effect on the more energetic carriers [see Fig.\ref{fig:jet_process}(d)]. 
A field of strength $B$ prevents carriers with energies below the threshold $\epsilon<eBvR/2$ from reaching the detectors at a distance $R$, providing a direct probe of the energy distribution of soft pairs.

Our system is described by the Hamiltonian for $N$ species of massless Dirac particles ($N = 4$ for graphene):
\begin{align}
\label{eq:formal_ham}
\HH &= \sum_{i= 1...N}\sum_{\vec{k}} \psi^{\dag}_{\vec{k}, i} \lp \hbar v \vec{\s} \cdot \vec{k}  \rp  \psi_{\vec{k}, i} + \HH_{\textrm{e-e}}
.
\end{align}
Here the optical field is included through minimal coupling $\vec{k} \to \vec{k}-\frac{e}{\hbar c}\vec{A}$ and $\HH_{\textrm{e-e}}$ describes e-e interactions \cite{Note1_bib}. We focus on the processes in a pristine material (undoped and disorder-free), assuming high mobility, long mean free paths and, for simplicity, ignoring the effects of electron-phonon scattering. While in real materials these effects may be significant, reducing the net response, they do not alter the outcome of competition between the on-shell and off-shell e-e processes. 

There are several ways to develop perturbation theory for e-e scattering: the weak-coupling approach uses small fine structure constant $\alpha=\frac{e^2}{\kappa \hbar v} \ll 1$, the large-$N$ approach uses as a small dimensionless coupling $1/N \ll 1$ with an RPA-screened interaction \cite{PhysRevB.59.R2474, PhysRevB.75.235423, PhysRevB.89.235431, PhysRevLett.113.105502}. The latter approach (which we use below) is in principle capable of dealing with systems at strong coupling $\alpha>1$ as long as the number of species $N$ is large enough. The resulting diagrammatics 
resembles that of QED, modulo replacing photon propagator by the dynamically screened Coulomb interaction \cite{PhysRevB.59.R2474}. 

\begin{figure}
\centering
\includegraphics[width=1.1\linewidth]{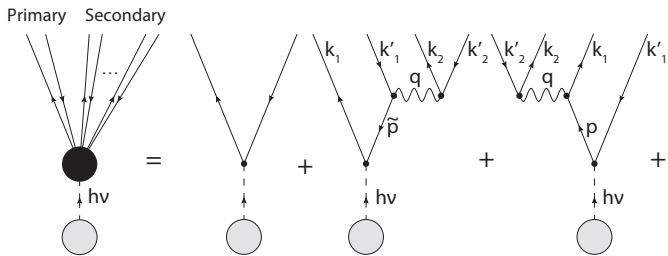}
\caption{Diagrammatic representation of single-photon absorption. Dashed lines describe interaction with a photon source, straight lines with arrows denote electron and hole propagators, wavy lines denote the dynamically screened Coulomb interaction, Eq.\eqref{eq:coulomb}.}
\label{fig:fey_diag}
\vspace{-5mm}
\end{figure}

A salient feature of Feynman diagrams describing the processes of secondary pair creation (see Fig.\ref{fig:fey_diag}) is the double-log divergences similar to those familiar in QED and QCD \cite{Gribov:1972ri, dokshitzer1991basics, Peskin:1995ev}. Below we analyze excitation of e-h pairs described by $\log^2$-divergent diagrams, which reflect production of infinitely many soft e-h pairs. We show that in the large-$N$ framework the rate for producing $p$ pairs behaves as $N^{-p} \log^{2p}$. Multiple $\log^2$ divergences can be tackled by 
resumming the contributions with the highest powers of $\log^2$\,\cite{Peskin:1995ev}, or by more refined approaches \cite{Gribov:1972ri, dokshitzer1991basics}. This approach allows us to obtain a detailed picture of the cascade, including the angular distribution and energy spectrum of secondary pairs.
We stress that the behavior of log divergences in graphene field theory is close to that in $(3+1)$-dimensional QED \cite{PhysRevB.59.R2474}, whereas the behavior in $(2+1)$-dimensional QED is quite different \cite{PhysRevD.23.2291, PhysRevD.41.1227}
but  is not directly relevant here.

We note that in a realistic setting the linear dispersion of Dirac bands, which is crucial for our analysis, is an asymptotic behavior valid at low enough energies. This makes the properties of soft pairs universal and largely insensitive to the details of band dispersion. 
For example, the trigonal warping is significant at high energies, but vanishes near the Dirac point \cite{RevModPhys.81.109}.
Another, potentially more critical, deformation of the Dirac cones arises due to interaction-induced velocity renormalization. The latter leads to dispersion `steepening' close to the Dirac point. This has two effects: one is further suppression of the on-shell relaxation rate, the other is a decrease in the phase-space  available for particles with small offshellness. However, since these effects occur at a first-log order, they are subleading to the $\log^2$ effects analyzed below. 

Photon absorption is represented diagrammatically as a sum of contributions with one incoming photon leg and many outgoing particle legs, with the screened e-e interaction replacing photon propagator in the corresponding QED diagrams. The lowest order tree-level diagrams are shown in Fig.\ref{fig:fey_diag}. The diagram with two particle legs describes creation of a primary e-h pair, an on-shell process with no virtual states. Such virtual states, present in the diagrams of higher order, are described by internal fermion lines without open ends. These states reside off shell, as indicated in Eq. \eqref{eq:offshellness_eq_1}. The higher-order diagrams describe creation of multiple secondary pairs, with summation over virtual states generating double-log divergences as discussed below. The wavy lines in Fig. \ref{fig:fey_diag} represent the dynamically screened interaction expressed through an exact polarization function as
\begin{align}
\label{eq:coulomb}
\tilde{V}_{\vec{q}, \omega} &= \frac{V_{\vec{q}}}{1 - V_{\vec{q}}\Pi(\vec{q},\omega)} 
, \quad 
V_{\vec{q}} =\frac{2\pi e^2}{\kappa |\vec{q}|}
,
\end{align}
with $\vec q$ and $\omega$ denoting the transferred momentum and frequency and
$\kappa$ is the dielectric constant. The values of $\kappa$ describing different substrates are discussed in [\onlinecite{Note1_bib}] 
along with the model used to generate Fig.\ref{fig:jet_process} 
and modification of $V_{\vec q}$ due to screening by the gate. 

Divergence in the polarization function $\Pi(\vec{q},\omega)$ 
softens the small-$q$ divergence of $V_{\vec{q}}$. We use a
simple expression \cite{PhysRevB.75.205418, PhysRevB.83.155441}, 
\be
\label{eq:pol}
\Pi(\vec{q},\omega) = -\frac{i N \vec{q}^2}{16\hbar}\frac{1}{\sqrt{\omega^2-v^2\vec{q}^2}}
,
\ee
describing the interband e-h pair excitations, $\omega>vq$. 

Crucially, even a single secondary pair creation is a strongly off-shell process. Indeed, linearity of band dispersion $\epsilon(\vec k)$ renders the e-e scattering processes obeying energy and momentum conservation to be of a strictly collinear character \cite{PhysRevLett.77.3589}. However, collinear scattering is subject to a phase space constraint that makes the transition rate vanish [see Fig.\ref{fig:jet_process}(a)]\cite{PhysRevB.83.155441}. In contrast, no phase space constraints arise for the off-shell processes [see Fig.\ref{fig:jet_process}(b)], and in fact the large phase space generates the double log-divergent contributions to the transition rate. 
This behavior extends to all higher-order multiple pair creation processes.

Turning to the quantitative analysis, we consider the second and third diagrams pictured in Fig.\ref{fig:fey_diag}, which describe an initial photoexcited e-h pair with energy and momentum positioned off-shell that excites a secondary e-h pair via an interband transition. At the end all participating particles are found in the on-shell states at the Dirac cone. The transition rate for this process, within the standard Golden Rule approach, takes the form: 
\be 
W_{0\to1}= \frac{2\pi}{\hbar} N^2
\!\!\!\!\!\!
\sum_{\vec{k}_1^\prime+\vec{k}_2^\prime = \vec{k}_1+\vec{k}_2}
\!\!\!\!\!\!
f_{\vec{k}_1^\prime}(1-f_{\vec{k}_1})f_{\vec{k}_2^\prime}(1-f_{\vec{k}_2}) \lvert \mathcal{A} \rvert^2\delta\lp \Sigma\epsilon_\alpha\rp
\label{eq:delta_sum_epsilon}
\ee 
Here $f_{\vec k}$ is the Fermi function, $h \nu$ is the absorbed photon energy (we set photon momentum equal zero), and $\delta\lp \Sigma\epsilon_\alpha\rp=\delta(\epsilon_{\vec{k}_1}+\epsilon_{\vec{k}_2}-\epsilon_{\vec{k}_1^\prime}-\epsilon_{\vec{k}_2^\prime}-h \nu)$. The transition matrix element $\mathcal{A}$ is given by a sum of two second-order contributions, which differ by the order of the operators describing photon absorption and secondary pair creation
\begin{align}
\label{eq:transition_matrix}
&\mathcal{A} = \la 
1,2 \lvert M_{\vec{q}, \omega} G(\epsilon_{\vec{p}}, \vec{p}) \vec{\sigma} 
\vec{A} + \vec{\sigma}  
\vec{A} G(\epsilon_{\tilde{\vec{p}}}, \tilde{\vec{p}}) M_{\vec{q}, \omega} \rvert 
1',2'\ra ,\nonumber\\
&\lvert M_{\vec{q}, \omega}\rvert^2 = \lvert\tilde{V}_{\vec{q}, \omega}\rvert^2 
\tilde F_{\vec{k}_2,\vec{k}_2^{\prime}} F_{\vec{k}_1, \vec{k}_1^{\prime}}
,
\end{align}
where $G(\epsilon, \vec{k})$ is the non-interacting fermion propagator,  and we introduced a shorthand notation $\lvert1,2\ra=\lvert\vec{k}_1, \vec{k}_2 \ra$, $\lvert 1',2'\ra=\lvert\vec{k}'_1, \vec{k}'_2 \ra$, using unprimed and primed symbols for the states of electrons and holes (see Fig.\ref{fig:fey_diag}). For brevity, we suppress the Dirac spinor structure and incorporate the factor $ve/c$ in the definition of the optical field $\vec{A}$ (to be restored below). 
The quantities $F_{\vec{k},\vec{k}^\prime}$ and $\tilde F_{\vec{k},\vec{k}^\prime}$ represent the coherence factors $\la \vec{k^\prime} s^{\prime} \rvert \vec{k} s\ra$ with $s=s'$ and $s\ne s'$,
describing the intraband and interband transitions, respectively \cite{doi:10.1143/JPSJ.75.074716}. The two terms in Eq.\eqref{eq:transition_matrix} describe the processes in which photon absorption is followed by a pair creation, and vice versa. The virtual states in the two contributions,  Eq.\eqref{eq:transition_matrix}, are characterized by the off-shell energy values: 
$\epsilon_{\vec p} = h \nu + \epsilon_{\vec{k}_1^{\prime}}$, $\vec{p} = \vec{k}_1^{\prime}$ and $\epsilon_{\tilde{\vec p}} = \epsilon_{\vec{k}_1}-h\nu$, $\tilde{\vec{p}} = \vec{k}_1$ (we use notations from Fig.\ref{fig:fey_diag}).

As will become clear shortly, the typical energy of secondary pairs $\omega$ is much smaller than the photoexcitation energy $h \nu$. Anticipating this result it is convenient to factorize the transition rate, expressing it through the spectral function of pair excitations. Following the standard route \cite{PhysRevB.26.4421} we first split the energy delta function in Eq.\eqref{eq:delta_sum_epsilon}:
\be 
\delta\lp \Sigma\epsilon_\alpha\rp =\int_{-\infty}^{\infty} \mathrm{d} \omega \delta(\epsilon_{\vec{k}_1}-\epsilon_{\vec{k}_1^\prime}-h \nu + \omega)\delta(\epsilon_{\vec{k}_2}-\epsilon_{\vec{k}_2^\prime}-\omega)\nonumber
\ee 
Next we use the identity $f_{\vec{k}^\prime}(1-f_{\vec{k}}) = (f_{\vec{k}^\prime}-f_{\vec{k}})(N_{\epsilon_{\vec{k}}-\epsilon_{\vec{k}^\prime}}+1)$, 
where $N_\omega=\frac1{e^{\beta\omega}-1}$ is the Bose function taken at the electron temperature, and rewrite the sum of $(f_{\vec k'_2}-f_{\vec k_2})\delta(\epsilon_{\vec k_2}-\epsilon_{\vec k'_2}-\omega)$ with the help of the relation
\be
\Im\Pi(\vec{q},\omega)=-N\pi\sum_{\vec k_2} \tilde{F}_{\vec k_2, \vec k'_2} (f_{\vec k'_2}-f_{\vec k_2})\delta(\epsilon_{\vec k_2}-\epsilon_{\vec k'_2} - \omega)\,,\nonumber
\ee
$\vec q = \vec k_2-\vec k'_2$, that follows from the definition of the polarization function \cite{PhysRevB.75.205418, 1367-2630-8-12-318}. This yields a more compact expression for the transition rate:
\begin{align}
\label{eq:W01_intermediate}
& W_{0\to1}= -\frac{2 N}{\hbar} \sum_{\vec{k}_1, \vec{k}_1^{\prime}, \vec{q}, \omega}
f_{\vec{k}_1^\prime}(1-f_{\vec{k}_1})(N_\omega 
+1)\lvert \mathcal{A}^{\prime}\rvert^2 
\\
&\times\Im\Pi(\vec{q},\omega) F_{\vec{k}_1,\vec{k}_1^{\prime}} 
| \tilde{V}_{\vec{q}, \omega}|^2 \delta_{\vec{k}_1^{\prime}, \vec{k}_1+\vec{q}}  \delta(\epsilon_{\vec{k}_1}-\epsilon_{\vec{k}_1^\prime}-h \nu + \omega) \nonumber
\end{align}
where $\omega$ and $\vec q$ are the
energy and momentum of the soft pair.  Here we introduced the quantity 
\be\label{eq:A'_one_pair}
\mathcal{A}^{\prime} = \la 1 \lvert  G(\epsilon_{\vec{p}}, \vec{p}) \vec{\sigma}\vec{A} + \vec{\sigma}\vec{A} G(\epsilon_{\vec{\tilde{\vec p}}}, \vec{\tilde{\vec p}})  \rvert 1^{\prime} \ra
\ee 
which represents the transition matrix element for the primary (`hard') pair, factoring out the contribution of the soft pair as described above (we again use a shorthand notation for the electron and hole states $\lvert \vec k_1\ra$ and $\lvert \vec k_1'\ra$ in Fig.1(c), for brevity suppressing the spin structure).

At this stage it is convenient to approximate the Green's functions of fermions in the virtual states [$G(\epsilon_{\vec p},\vec p)$ and $G(\epsilon_{\vec{\tilde{\vec p}}}, \vec{\tilde{\vec p}})$ in Eq.\eqref{eq:A'_one_pair}] by expanding in the small frequency $\omega$ and momentum $\vec q$ transferred to the soft pair. 
This is done by writing $\epsilon_{\vec{p}} = \epsilon_{\vec{k}_1} + \omega$, $\vec{p} = \vec{k}_1 + \vec{q}$ and $\epsilon_{\vec{\tilde{\vec p}}} = \epsilon_{\vec{k}_1^{\prime}}-\omega$, $\tilde{\vec{p}} = \vec{k}_1^{\prime}-\vec{q}$ and expanding in
$\omega$ and $\vec q$.
The approximation that uses the softness of the secondary pair as a small parameter is known as the `eikonal approximation', since at small $\omega$ and $\vec q$ only the phase of the fermion wavefunction varies but not the spinor part. 
Suppressing the spinor part, we obtain simple expressions
\begin{align}\label{eq:eikonal}
G(\epsilon_{\vec{p}}, \vec{p}) \approx \frac{-1}{\omega+v q_{\parallel}}
,\quad G(\epsilon_{\tilde{\vec{p}}}, \tilde{\vec{p}}) \approx \frac{1}{\omega-v q_{\parallel}},
\end{align}
where $q_{\parallel}$ is the component of $\vec{q}$ parallel to $\vec{k}_{1}$. The two terms in Eq.\eqref{eq:eikonal} originate from the corresponding electron and hole contributions in Eq.\eqref{eq:A'_one_pair}. We note parenthetically that the denominators in Eq.\eqref{eq:eikonal} do not vanish since the soft pairs obey $|\omega|>v|\vec q|$. The matrix element $\mathcal{A}'$ is then reduced to
\be
\mathcal{A}^{\prime} \approx \frac{2 v q_{\parallel} \la 1 | \vec{\sigma}\vec{A} | 1^{\prime} \ra}{\omega^2-v^2 q_{\parallel}^2}\,
.
\ee
After plugging it in Eq.\eqref{eq:W01_intermediate}, the quantity $W_{0\to1}$ becomes
\be
\label{eq:perturbation_penultimate}
W_{0\to1} = -\frac{8 N}{\hbar} \sum_{\vec{k}_1, \vec{q}}
|\tilde{V}_{\vec{q}, \omega}|^2 
\Im\Pi(\vec{q},\omega) \left\lvert\frac{v q_{\parallel} \la 1 | \vec{\sigma}\vec{A} | 1' \ra}{\omega^2-v^2 q_{\parallel}^2}\right\rvert^2
,
\ee
where $\omega = h \nu - 2 v \lvert \vec{k}_1\rvert - v q_{\parallel}$. To arrive at Eq.\eqref{eq:perturbation_penultimate} we approximated the intraband  coherence factor by unity, since $F_{\vec k_1, \vec k_1 + \vec q} \approx 1$ in the soft-pair limit $q\ll k_1$. 
The interband coherence factor $F$ has been included in the soft pair spectral function through the factorization procedure outlined above. The factor $N_\omega + 1$, which we suppressed for brevity, limits summation in Eq.\eqref{eq:perturbation_penultimate} to $\omega>0$ for $T=0$. At $T>0$, somewhat counterintuitively, this factor does not impact or regulate the IR divergence (see Refs. [\onlinecite{Note1_bib}] and [\onlinecite{lewandowski2017}] for detailed discussion).

The 
transition rate $W_{0\to1}$ 
features a
double-log divergence originating from the collinear 
e-e scattering. The divergence arises due singular behavior of the quantities in Eq.\eqref{eq:perturbation_penultimate} upon integration upon the soft-pair momentum $\vec q$. In that, one log divergence arises from the integral over the length $|\vec q|$, the other log comes from integration over the angle between $\vec q$ and $\vec k_1$. For a quantitative estimate we evaluate the double-log contribution at leading order in $1/N$, which can be done by approximating $\tilde{V}_{\vec{q}, \omega}\approx -1/\Pi(\vec{q},\omega)$. After integrating over $\vec q$ and $\vec k_1$, and factoring out $W_{\text{on-shell}}$, the transition rate for the on-shell diagram in Fig.\ref{fig:jet_process}(e), 
the 
rate $W_{0\to1}$ becomes
\be
\label{eq:w_gr}
\frac{W_{0\to1}}{W_{\text{on-shell}}} \approx  \frac{8}{N \pi^2} \lp \ln \frac{\epsilon_{>}}{\epsilon_{<}}\rp^2
,\quad 
W_{\text{on-shell}} = \frac{e^2 \vec{A}^2 h \nu}{c^2} \frac{N}{8}
,
\ee
where $\approx$ indicates that we contributions subleading to double log were suppressed \cite{lewandowski2017, Note1_bib}. Here the UV cutoff  $\epsilon_{>}$ is of order $h \nu/2$ (energy of an excited electron immediately after photon absorption). The IR cutoff $\epsilon_{<}$ is set by the Dirac point width, controlled by carrier collisions or disorder.
The $\log^2$ divergence in Eq.\eqref{eq:w_gr} is a direct consequence of linear dispersion, arising from soft secondary pairs 
that are near-collinear with respect to the primary pair direction and form two counterpropagating jets.

The double-log divergence in the transition rate is reminiscent of the double-log divergences familiar from QCD or QED calculations. This can be seen e.g. by comparing to soft Bremsstrahlung in QED \cite{Peskin:1995ev}, and noting that the double logs arise in an identical manner in both cases, with one log originating from an integral over momentum magnitude and the other from angular integration. As in QED, the IR double-log divergence means that the secondary pairs are much softer than the primary pair, vindicating our eikonal approximation.

The jets formed by soft pairs have random spatial orientation, aligned with the e and h velocities of parent hard pairs [see Fig.\ref{fig:jet_process}(c)-(e)].  The mean number of pairs in a jet is estimated below. 
Each jet
features a sharp angular distribution that peaks at $\theta=0$, $\pi$ relative to the parent pair direction. The corresponding counting distribution, normalized to the total number of secondary pairs (see [\onlinecite{Note1_bib}]), is shown in Fig.\ref{fig:jet_process}(e). Energy distribution of soft pairs has a power-law tail at low energies \cite{lewandowski2017}.

We parenthetically note that dynamical screening, Eq.\eqref{eq:coulomb}, is crucial for our analysis. Had an unscreened Coulomb interaction $V_{\vec q}$ been used, the transition rate would have been IR divergent as a power law rather than as $\log^2$. This is in line with the argument that the perturbation series for Dirac semimetals should be carried out in powers of a screened interaction rather than the bare one \cite{PhysRevLett.113.105502}. This behavior is in contrast to QED, where double-log divergences arise from perturbation theory in bare coupling.

Motivated by the resemblance to QED, the higher-order contributions of the form $N^{-n}\log^{2n}$ can be analyzed by a Sudakov-like resummation scheme of leading double-log divergent diagrams. These diagrams describe primary pair creation followed by emission of multiple secondary pairs in analogy to `hard' scattering processes in QED accompanied by emission of soft photons. There are soft e-h pairs of two distinct types emitted, respectively, by the hard electron and the hard hole. These soft pairs form two counterpropagating jets [see Fig.\ref{fig:jet_process}(c)-(e)]. For each of the two jets, in the limit of 
the emitted pairs being independent of one another and assuming 
no mutual phase-space blocking, the probability distribution is Poissonian \cite{Peskin:1995ev},  
\be
p_n=\frac{{\tilde\lambda}^n}{n!} e^{-\tilde \lambda}
,\quad
\tilde\lambda= \frac{4}{N \pi^2} 
\lp \ln \frac{\epsilon_{>}}{\epsilon_{<}}\rp^2
.
\ee 
The value $\tilde\lambda$ is a half of the total single-pair emission rate given in Eq.\eqref{eq:w_gr}. Combining two identical Poisson distributions gives a Poisson  counting distribution with a double rate accounting for both jets \cite{Note1_bib}: 
\be
\label{eq:transition_rate_n_pairs}
\frac{W_{0\to n}}{W_{\text{on-shell}}} = \frac{\lambda^n
e^{-\lambda}}{n!}
,\quad
\lambda = 2\tilde\lambda = \frac{8}{N \pi^2} \lp\ln \frac{\epsilon_{>}}{\epsilon_{<}}\rp^2
.
\ee
The mean number of secondary pairs $\langle N_{\text{sec}} \rangle=\lambda$ 
goes as $\log^2$ and hence can be much greater than unity. As an illustration, a $h\nu = 1\,{\rm eV}$ photon creates
between $4$ and $10$ pairs for ratios $\epsilon_>/\epsilon_< = 10^2$--$10^3$,
which corresponds
to realistic
Dirac point widths.

Interestingly, the process in which no soft pairs are emitted has a vanishing rate. Indeed, $W_{0\to0}$ vanishes in the limit $\epsilon_{<}\to 0$. To interpret this result we note that the sum of all partial rates equals the bare on-shell rate:  
$\sum_{n=0}^\infty W_{0\to n}=W_{\text{on-shell}}$. This means that massive  emission of soft pairs does not alter the net photon absorption probability. 
Instead, the absorbed photon energy is 
redistributed among a large number of secondary e-h pairs, providing a mechanism for carrier multiplication. 

In summary, the off-shell pathways unblock kinematic constraints for collinear scattering in a Dirac band, allowing a large number of secondary pairs to be produced as the photogenerated carriers cascade down in energy. The angular distribution of secondary pairs is sharply peaked along the primary pair velocity, representing a condensed-matter analog of relativistic jets familiar from high-energy physics. 
The jets can be directly probed using a solid-state equivalent of particle detectors as discussed above [Fig.\ref{fig:jet_process}(c),(d)]. Formation of jets is corroborated by recent experimental studies of Auger scattering processes \cite{PhysRevLett.117.087401, PhysRevB.94.235430}, which indicate that at weak electron-phonon coupling the collinear scattering processes dominate the relaxation pathways of photoexcited carriers.

\begin{acknowledgments}
We acknowledge support of the Center for Integrated Quantum Materials under NSF award DMR-1231319, and MIT Center for Excitonics, an EFRC funded by the U.S. Department of Energy,
Office of Science, Basic Energy Sciences under Award no. DE-SC0001088.
\end{acknowledgments}

\pagebreak
\widetext
\begin{center}
\textbf{\large Supplemental Material for ``Photoexcitation Cascade and Quantum-Relativistic Jets in Graphene"}
\end{center}
\setcounter{equation}{0}
\setcounter{figure}{0}
\setcounter{table}{0}
\setcounter{page}{1}
\makeatletter
\renewcommand{\theequation}{S\arabic{equation}}
\renewcommand{\thefigure}{S\arabic{figure}}

\section{The structure of the Hamiltonian}
\label{sec:structure_hamil}

Electrons in graphene are described by the Hamiltonian for $N$ species of massless Dirac particles:
\be
\HH = \sum_{\vec{k}, i} \psi^{\dag}_{\vec{k}, i} \lb v \vec{\s} \cdot \lp\hbar\vec{k} - \frac{e}{c} \vec{A}( \vec{r}, t ) \rp \rb  \psi_{\vec{k}, i} +  \frac{1}{2} \sum_{\vec{q}, \vec{k}, \vec{k}^{\prime}, i, j} V_{\vec{q}} \psi^{\dag}_{\vec{k}+\vec{q}, i} \psi^{\dag}_{\vec{k}^\prime-\vec{q}, j} \psi_{\vec{k}^\prime, j} \psi_{\vec{k}, i} .
\ee
Here $i,j = 1\dots N$ and $N=4$ is the spin/valley degeneracy, $\psi_{\vec{k}, i}$, $\psi^{\dag}_{\vec{k}, i}$ describe two-component Dirac fermions, and in the last term we suppressed the inner products of $\psi$ and $\psi^\dag$. For pristine free-standing graphene the carrier-carrier interaction is the unscreened Coulomb $1/r$ interaction, giving $V_{\vec{q}} = \frac{2\pi e^2}{|\vec{q}|}$. The vector potential $\vec{A}( \vec{r}, t )$, linearly coupled to the current operator, describes the optical field. Since optical wavelengths are large compared to the characteristic wavelengths of photoexcited carriers, we ignore the $\vec r$ dependence in $\vec{A}( \vec{r}, t )$, treating it as a spatially uniform time-dependent perturbation.

The effect of the substrate is accounted for by a mean dielectric constant as  
\be\label{eq:mean_kappa}
V_{\vec{q}} = \frac{2\pi e^2}{|\vec{q}|\kappa}
,\quad
\kappa =  \frac{\kappa_1+\kappa_2}2
.
\ee 
Here $\kappa_1$ and $\kappa_2$ are the bulk permittivity values of the material above and below the graphene sheet. We assume a dielectric on one side of the sheet and vacuum or
air ($\kappa_{\textrm{air}}=1$) on the other side. E.g. for graphene on hBN substrate, using $\kappa_{\textrm{hBN}}=5.06$, gives the mean dielectric constant value  $\kappa_{\textrm{G/hBN}}=(\kappa_{\textrm{air}}+\kappa_{\textrm{hBN}})/2=3.03$.
Likewise, for SiC substrate the bulk value $\kappa_{\textrm{SiC}} = 10.04$ yields the 
mean value 
$\kappa_{\textrm{G/SiC}}=5.52$. These values are used to generate the curves shown in Fig.\ref{fig:jet_process} of the main text and in Fig.\ref{fig:app_N_vs_g}.  

While the main text focuses on the interaction given in Eq. \eqref{eq:mean_kappa}, we also considered the effect of screening by a gate placed a distance $H$ below the graphene sheet. The change of potential in the presence of the gate can be accounted for by image charges at a distance $2H$ beneath graphene plane, which modifies the interaction as 
\be\label{eq:V_gate}
V_{\vec{q}}=\frac{4\pi e^2}{|\vec q| \lp \kappa_1+\kappa_2 \coth(|\vec q|H)\rp }
\ee
where $\kappa_{1,2}$ are the dielectric constant values specified above. This expression matches the one in Eq. \eqref{eq:mean_kappa} when the gate is removed (i.e. in the limit $H\to \infty$). The gate screens out the long-wavelength harmonics with $q\lesssim \frac1{2H}$ introducing a new energy scale $\hbar v/2H$. Here we consider the effect of the gate only on the angular distribution of the counts, the effect of the gate on the number of pairs and transition rate is discussed elsewhere.

In our analysis we assume constant, frequency-independent permittivity values $\kappa_{1,2}$ and treat the gate as an ideal conductor. The dynamical response of the electron gas is included via the polarization function $\Pi(\vec{q},\omega)$ given in Eq. \eqref{eq:pol}. 
The results for gated graphene are shown in Fig.\ref{fig:jet_process} of the main text and in Fig.\ref{fig:app_N_vs_g}.

The dependence in Eqs.  \eqref{eq:V_gate} and   \eqref{eq:mean_kappa} can be derived using Fourier expansion of the 3D potential $\phi(\vec r)$ of a point charge $e$ placed at the graphene plane
\be
\phi(\vec r)=\sum_{\vec q} e^{i\vec q\vec r_\parallel}\phi_{\vec q}(z)
\ee
where $\vec q$ is a two-dimensional wavevector and $\vec r_\parallel$ denotes the radius vector component parallel to graphene plane. Potential $\phi(\vec r)$ satisfies Laplace's equation in 3D, which yields an ordinary differential equation for the Fourier coefficients in the two regions $z>0$ and $-H<z<0$ above and below graphene plane:
\be
(\p_z^2-\vec q^2)\phi_{\vec q}(z)=0
.
\ee
At $z=0$ the Fourier coefficients obey a matching condition derived from Gauss' law,  
\be\label{eq:Gauss_law}
-\kappa_1\p_z\phi_{\vec q}(z=0+)+\kappa_2\p_z\phi_{\vec q}(z=0-)=4\pi e
,
\ee
and a continuity condition. Taking a solution that vanishes at $z=-H$ and decays exponentially at $z\to \infty$ we have 
\be
\phi_{\vec q}(z>0)=V_{\vec q} e^{-|\vec q| z}
,\quad
\phi_{\vec q}(-H<z<0)=V_{\vec q} \frac{\sinh |\vec q| (z+H)}{\sinh |\vec q| H}
.
\ee
The value $V_{\vec q}=\phi_{\vec q}(z=0)$ can then be determined by plugging this dependence in Eq.  \eqref{eq:Gauss_law}, which
gives the result in Eq. \eqref{eq:V_gate}. Taking the limit $H\to\infty$ gives Eq. \eqref{eq:mean_kappa}.

\section{Poissonian counting distribution for one and two jets}

There is a simple relation between contributions to the counting statistics from the processes involving $n=1$ and $n>1$ emitted soft pairs. The former are described by the low-order Feynman diagrams discussed in the main text, whereas the latter are described by higher-order diagrams accounting for 
primary pair creation followed by emission of multiple soft secondary pairs. These soft secondary pairs are created either by the hard electron or the hard hole. 
The probability distributions $p_n^{e,h}$ describing the numbers of such pairs $n$ emitted by a given parent hard particle, e or h, have simple properties in the limit when the emitted pairs can be treated as being independent of one another. This is the case at weak coupling when the emitted pairs do not interact with one another. In this case   
we expect the probability distributions $p_n^{e,h}$ to be Poissonian. Here we demonstrate that two identical Poisson distributions $p_n^{e,h}$, when combined together, give rise to a Poisson distribution with a double rate. As defined in the main text:
\be
p_n^{e,h}=\frac{{\tilde\lambda}^n}{n!} e^{-\tilde \lambda}
,\quad
\tilde\lambda= \frac{\lambda}{2}=\frac{4}{N \pi^2} 
\lp \ln \frac{\epsilon_{>}}{\epsilon_{<}}\rp^2
.
\ee 
Probability of emitting $n$ soft eh pairs from either the hard electron or the hard hole is thus given by
\be
P_n = \sum_{m=0 } ^n 
p_m^e p_{n-m}^h
\ee
Plugging $p_n^{e,h}$ gives a Poisson distribution with the double rate $\lambda=2\tilde\lambda$, see Eq.\eqref{eq:transition_rate_n_pairs} in the main text:
\be
P_n=\frac{\lambda^n}{2^n}e^{-\lambda}\sum_{m=0 } ^n
\frac{1}{m!(n-m)!}=\frac{\lambda^n e^{-\lambda}}{n!}
.
\ee
Here the sum over $m$ is evaluated using 
the binomial formula $\sum_{m=0 } ^n \bigl( \begin{smallmatrix} n \\ m\end{smallmatrix}\bigr) = 2^n$. 

The meaning of the double rate $\lambda$ is that the soft pairs emitted 
following the primary photoexcitation event by both hard particles, e and h, 
appear nearly simultaneously but are statistically uncorrelated. It should be noted, however, that 
the soft pairs emitted by each parent particle have different spatial structure, forming two counterpropagating jets directed along the e and h velocities. The counting distribution in each of the two jets is Poissonian with the half rate $\tilde\lambda=\lambda/2$.

\section{The angular distribution of the counting rate}
\label{sec:ang_dis}
Here we discuss the angular distribution of secondary pairs in a system with and without an applied gate. Some of the initial analysis is analogous to the derivation presented in Ref.[\onlinecite{lewandowski2017}], which we restate here for reader's convenience.

The angular distribution of secondary pairs $p(\theta)$ can be obtained from the expression for the transition rate $W_{0\to1}$ (Eq.\eqref{eq:perturbation_penultimate} in the main text),
\be
\label{eq:pen_ult_step_ang}
W_{0\to1} = -\frac{8 N}{\hbar} \sum_{\vec{k}_1, \vec{q}} ( N_\omega +1)
|\tilde{V}_{\vec{q}, \omega}|^2
\Im\Pi(\vec{q}, \omega) \left\lvert\frac{v q_{\parallel} \la 1 | \frac{ev}{c} \vec{\sigma}\vec{A} | 1' \ra}{\omega^2-v^2 q_{\parallel}^2}\right\rvert^2
,
\ee
where $\vec k_1$ is the momentum at which the primary hard pair is excited, $\vec q$ is the momentum transferred to the soft pair and $\omega = h \nu - 2 v \lvert \vec{k}_1\rvert - v q_{\parallel}$ is the energy of the soft pair. We define $p(\theta)$ by factorizing Eq. \eqref{eq:pen_ult_step_ang} as
\be
\label{eq:def_ang}
W_{0\to1}= W_{\text{on-shell}} \int_0^{2\pi} d\theta ~p(\theta)  ,
\ee
where $W_{\text{on-shell}}$ is the transition rate for emission of only the primary pair and
\be
\label{eq:ang_dis_pen}
p(\theta) =-\frac{8 N}{(2\pi)^2 \hbar} \frac{1}{W_{\text{on-shell}} } \sum_{\vec{k}_1} \int_0^{\infty} q d q  \left(N_\omega+1\right)
|\tilde{V}_{\vec{q}, \omega}|^2
\Im\Pi(\vec{q}, \omega) \left\lvert\frac{v q \cos\theta \la 1 | \tfrac{ev}{c} \vec{\sigma}\vec{A} | 1' \ra}{\omega^2-v^2 q^2 \cos^2\theta}\right\rvert^2\,.
\ee
Here we introduced a polar coordinate system
\be
q_{\parallel} = q \cos\theta ,\quad q_{\perp} = q \sin\theta\,,
\ee
with $q = |\vec{q}|$ and $\theta$ being the angle between $\vec{q}$ and $\vec{k}_1$.

The analysis is facilitated by expressing the integral over $\vec k_1$ through an integral over the soft-pair frequency $\omega=h\nu-2v|\vec k_1|-v q \cos\theta$. 
This is done by writing the sum over $\vec{k}_1$ as
\be
\label{eq:k1}
\sum_{\vec{k}_1} \dots = \int_{0}^{\infty} \frac{d{} | \vec{k}_1 |}{2\pi} | \vec{k}_1| \int_{0}^{2\pi} \frac{d{} \theta_{\vec k_1}}{2\pi} \dots \approx \frac{h \nu}{8\pi v^2} \int_{-\infty}^{\infty} 
d\omega \int_{0}^{2\pi}\frac{d{} \theta_{\vec{k}_1}}{2\pi} \dots
\ee
where we used the soft-pair approximation $\omega\ll h\nu$ to introduce a constant density of states at half the photon energy $\epsilon=h\nu/2$ (Eq. (14) in [\onlinecite{lewandowski2017}]). Inserting Eq.\eqref{eq:k1} into Eq.\eqref{eq:ang_dis_pen} we note that the dependence on $\vec k_1$ orientation relative to $\vec A$, i.e. on the angle $\theta_{\vec k_1}-\theta_{\vec A}$, is present only in the matrix element $\la 1 | \tfrac{ev}{c} \vec{\sigma}\vec{A} | 1' \ra$. We can therefore carry out the integration over $\theta_{\vec k_1}$ as
\be
\label{eq:k_1_av}
\int_{0}^{2\pi}\frac{d{} \theta_{\vec{k}_1}}{2\pi} 
\Big| \langle 1 \lvert \textstyle{\frac{ev}{c}} \vec{\sigma} \vec{A} \rvert 1' \rangle \Big|^2
= \frac{e^2 v^2 \vec{A}^2}{2 c^2}
\,.
\ee
With this simplification we can rewrite the angular distribution as
\be
\label{eq:app_angular}
p(\theta) = \frac{-1}{\pi^3 \hbar} \int_{-\infty}^{\infty} d\omega~\int_{0}^{\infty} d q ~ q \left(N_\omega+1\right)
|\tilde{V}_{\vec{q}, \omega}|^2
\Im\Pi(\vec{q}, \omega) \frac{v^2 q^2 \cos^2\theta}{(\omega^2-v^2 q^2 \cos^2\theta)^2}\,,
\ee
where we used the expression for on-shell transition rate $W_{\text{on-shell}}=\frac{e^2\vec A^2 h\nu}{c^2}\frac{N}8$ (Eq. (7) in [\onlinecite{lewandowski2017}]).

To systematically account for the IR divergences we employ the same regularisation scheme as in [\onlinecite{lewandowski2017}] specifically the non-zero mass polarization operator:
\be
\Pi(\vec{q},\omega)=-\frac{iN\vec q^2}{16\hbar v}\frac1{\sqrt{(\omega/v)^2-\vec q^2-k_0^2}}\,.
\ee
 The angular distribution is therefore
\be
p(\theta) = \frac{16}{N\pi^3} \int_{-\infty}^{\infty} d\left(\frac{\omega}{v}\right) \int_0^{\infty} d q \frac{g^2 q \left(N_\omega+1\right)}{q^2 \left(1+g^2\frac{q^2}{\frac{\omega^2}{v^2}-q^2-k_0^2}\right)} \frac{q^2 \Theta\left(\frac{\omega^2}{v^2}-q^2-k_0^2\right)}{\sqrt{\frac{\omega^2}{v^2}-q^2-k_0^2}} \frac{q^2 \cos^2\theta}{\left(\frac{\omega^2}{v^2}-q^2 \cos^2\theta\right)^2},
\ee
where we introduced a dimensionless coupling constant $g = \pi N \alpha/8$. The origin of each term can be found by comparison with Eq.  \eqref{eq:app_angular}. 
Rationalizing the expression and rescaling $\omega/v \to \omega$ we arrive at:
\be
\label{eq:final_ang}
p(\theta) = \frac{16}{N\pi^3} \int_{0}^{\infty} d \omega \int_0^{\infty} d q \frac{g^2 q^3 \cos^2\theta \sqrt{\omega^2-q^2-k_0^2} ~\Theta\left(\omega^2-q^2-k_0^2\right)}{\left(\omega^2-k_0^2+(g^2-1)q^2\right)\left(\omega^2-q^2 \cos^2\theta\right)^2}\,.
\ee
We replaced the integral $\int_{-\infty}^{\infty} d \omega ~(N_\omega+1)$ with $\int_0^{\infty} d \omega$, which follows from the identity $N_\omega +N_{-\omega}+1=0$ as
\be
\int_{-\infty}^\infty (N_\omega +1) F(\omega) d\omega
= \int_0^{\infty} F(\omega) d\omega
\ee
is valid for any even integrable function $F(\omega)$ (Eq. (16) in [\onlinecite{lewandowski2017}]). Note that this implies that dependence on temperature disappears from the angular distribution calculation and does not regularize the infrared divergence of Eq. \eqref{eq:final_ang}.
 
\begin{figure}
\includegraphics[width=0.75\linewidth]{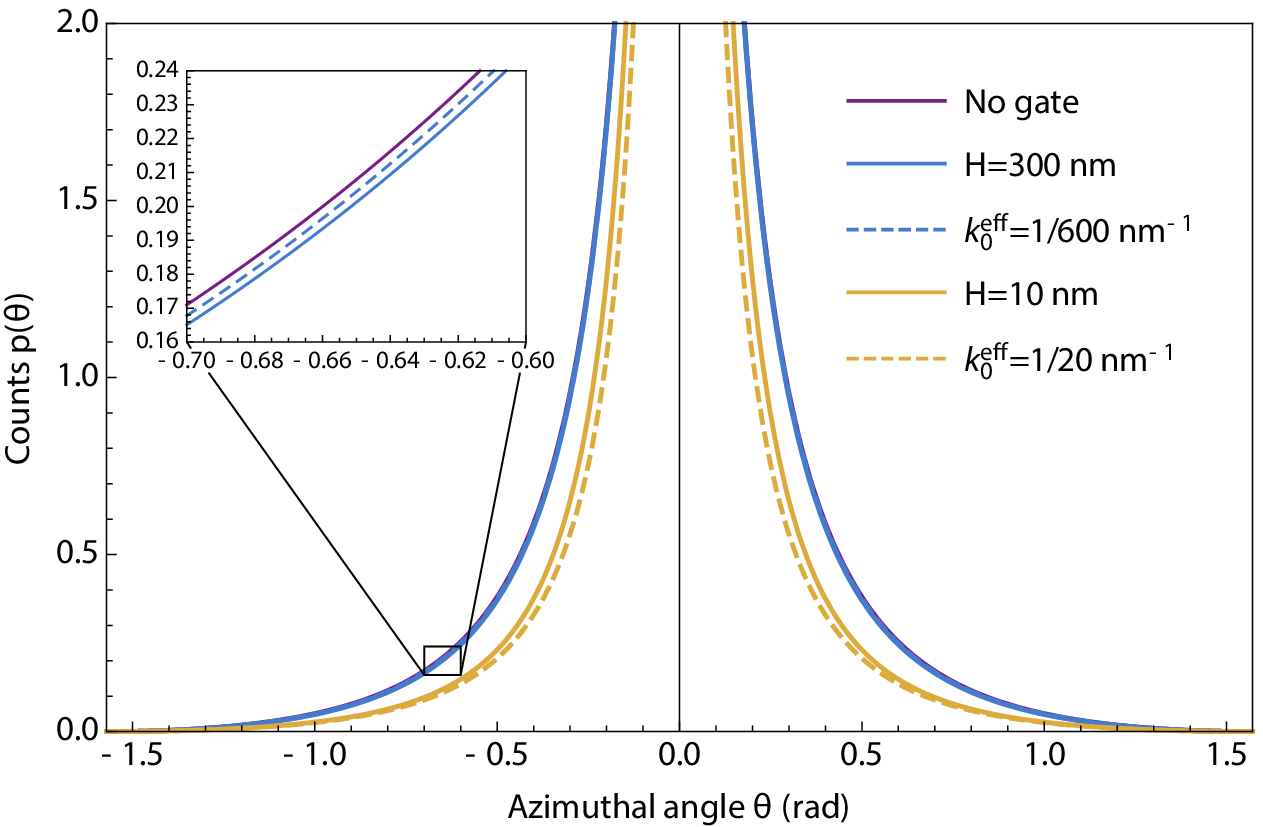}
\caption{The angular distribution of counts $p(\theta)$ for a system with and without the gate. The effect of the gate is to screen out long wavelength harmonics $q \to 0$ leading to a narrower counts distribution. The curves for $k_0^{\rm eff}$ correspond to a system without a gate, but with a lower limit of the $q$ integral in Eq. \eqref{eq:final_ang} replaced with $k_0^{\rm eff}=\frac1{2H}$ as explained in text. A gate at $H=300\,{\rm nm}$ has almost no effect on the angular distribution as the effective $q$ integral cutoff is comparable to the $k_0$ one (using $k_\nu/k_0 = 10^3$ for a $1\,{\rm eV}$ photon).}
\label{fig:app_N_vs_g}
\end{figure}

With the non-zero mass regularisation the angular distribution $p(\theta)$ is IR safe, however it still has a UV divergence in the integration over $\omega$. To control it we replace the upper limit of the $\omega$ integral as
\begin{align}
\label{eq:control_uv_ir}
\int_{0}^{\infty} d{} \omega  &\to \int_{0}^{k_\nu} d{} \omega \,,
\end{align} 
where $k_\nu = h\nu /2\hbar v$ is the photon's energy. In Fig. \ref{fig:jet_process}(e) and Fig. \ref{fig:app_N_vs_g} we plot the angular distribution $p(\theta)$ numerically integrated for the ratio $\epsilon_>/\epsilon_< = k_\nu/k_0 = 10^3$. We note that the effect of varying the substrate does not significantly alter the angular's distribution shape.

The plotted angular distribution $p(\theta)$ has an apparent divergence near $\theta \to 0$ (and $\theta\to\pi$). To see that analytically, we take the large-$N$ limit ($g\to \infty$) and rewrite the equation Eq.\eqref{eq:final_ang} as
\be
\label{eq:p_theta_small_1}
p(\theta) = \frac{8}{N\pi^3} \int_{0}^{\infty} d \omega \int_0^{\infty} d (q^2) \frac{\cos^2\theta \sqrt{\omega^2-q^2-k_0^2} ~\Theta\left(\omega^2-q^2-k_0^2\right)}{\left(\omega^2-q^2 \cos^2\theta\right)^2}\,,
\ee
where we also note the change of variables from $q \to q^2$.
With the help of an integral (for $b > a$):
\be
\int_0^{a^2} d x \frac{\sqrt{a^2-x}}{(b^2-x)^2} = -\frac{a}{b^2}+\frac{\sin^{-1}\frac{a}{b}}{\sqrt{b^2-a^2}}\,,
\ee
we carry out the integration over $q^2$ and arrive at
\be
p(\theta) =\frac{8}{N\pi^3} \int_{k_0}^{k_\nu} d \omega \left(\frac{\sin^{-1}\left(|\cos\theta|\sqrt{1-\frac{k_0^2}{\omega^2}}\right)}{\omega|\cos\theta|\sqrt{\sin^2\theta+\frac{k_0^2}{\omega^2}\cos^2\theta}}-\frac{\sqrt{1-\frac{k_0^2}{\omega^2}}}{\omega}\right)\,,\label{eq:ang_large_n}
\ee
which is valid at all angles in the large-$N$ approximation.
Focusing only on the $\theta$ dependent part near $\theta=0$ (in the limit of $k_0/\omega  \ll 1$) we get
\be
\label{eq:ang_dis_no_gate}
p(\theta) \approx \frac{8}{N\pi^3} \int_{k_0}^{k_\nu} d \omega \frac{1}{\sqrt{k_0^2+\omega^2\theta^2}} = \frac{1}{\theta}\sinh^{-1}\frac{k_\nu \theta}{k_0}
\ee
where we kept only the term controlled by the ratio $k_\nu \theta/k_0$. For $\theta > \frac{k_0}{k_\nu}$ the angular distribution $p(\theta)\propto \frac{1}{\theta} \ln \frac{2 k_\nu \theta}{k_0}$, however as $\theta < \frac{k_0}{k_\nu}$ the divergence is regularised to $p(\theta)\propto \frac{k_\nu}{k_0}$ and becomes analytic in $\theta$.

In a similar manner we can investigate the effect of a gate located at a distance $H$ from the graphene substrate on the distribution of the generated pairs. As described previously, the presence of a gate leads to formation of image charges at a distance $2H$ away from graphene. This is captured by modifying the Fourier transform of the bare Coulomb interaction $V_{\vec{q}}$ as in Eq. \eqref{eq:V_gate} and repeating the calculation from Eqs. \eqref{eq:app_angular}-\eqref{eq:final_ang}. A quick way of arriving at the same result is to note that each $V_{\vec{q}}$ term essentially gives rise to one $g$ coupling constant factor. We can therefore perform the following substitution
\be
\label{eq:gate_mod}
g^2 \to \frac{g^2}{\left(\frac{\kappa_1}{2\kappa}+\frac{\kappa_2}{2\kappa}\coth(qH)\right)^2}
\ee
in the expression Eq.  \eqref{eq:final_ang} above. Here $\kappa=(\kappa_1+\kappa_2)/2$ is the mean permittivity for the same geometry, but without the gate.

The introduction of a gate at a distance $H$ will lead to 
screening of the $1/r$ interaction at distances greater than $\sim2H$. This translates into a new IR energy scale, which is competing with the one set by mass $k_0$ as $k_0^{\rm eff} \approx {\rm max} \left(k_0, 1/2 H\right)$ \cite{lewandowski2017}. To better understand this we go back to the analysis of Eq.\eqref{eq:final_ang}, 
\be
\label{eq:final_ang_gate}
p(\theta) = \frac{16}{N\pi^3} \int_{0}^{\infty} d \omega \int_0^{\infty} d q \frac{g(q)^2 q^3 \cos^2\theta \sqrt{\omega^2-q^2-k_0^2} ~\Theta\left(\omega^2-q^2-k_0^2\right)}{\left(\omega^2-k_0^2+(g(q)^2-1)q^2\right)\left(\omega^2-q^2 \cos^2\theta\right)^2}\,.
\ee
where the dimensionless constant $g$ became momentum $q$ dependent to account for presence of a gate (as explained earlier with reference to Eq.\eqref{eq:gate_mod}). In order not to obscure the analysis with unnecessary constants, we change our graphene setup from air/graphene/dielectric to a case of dielectric/graphene/dielectric geometry. This implies that $\kappa_1 = \kappa_2 = \kappa$ in Eq.\eqref{eq:gate_mod} and hence
\be
\label{eq:gate_mod_simp}
g(q) = g_0(1-e^{-2Hq})\,.
\ee
Here $g_0 = \pi N \alpha/8$ is the dimensionless coupling constant.  The gate regularizes our integral at low $q$ and hence we can take $k_0 \to 0$. With these provisions the angular distribution $p(\theta)$ from Eq. \eqref{eq:final_ang_gate} becomes:
\be
p(\theta) = \frac{16}{N\pi^3} \int_{0}^{\infty} d \omega \int_0^{\infty} d q \frac{g(q)^2 q^3 \cos^2\theta \sqrt{\omega^2-q^2} ~\Theta\left(\omega^2-q^2\right)}{\left(\omega^2+(g(q)^2-1)q^2\right)\left(\omega^2-q^2 \cos^2\theta\right)^2}\,.
\ee
To satisfy the Heaviside function $\Theta\left(\omega^2-q^2\right)$ we perform a variable change $\omega = \sqrt{y+1}q$:
\be
p(\theta) =  \frac{8}{N\pi^3} \int_0^{\infty} d q \frac{g(q)^2 \cos^2\theta}{q} \int_0^{\infty} d y \frac{\sqrt{y}}{\sqrt{y+1}(y+g(q)^2)(y+1-\cos^2\theta)^2}\,.
\ee
From the structure of the polarization operator we know that the region $\omega \approx q$ gives rise to the IR divergences. We focus therefore on the $y\to 0$ region (taking $\sqrt{y+1}\approx 1$ for $y \to 0$):
\be
p(\theta) \approx \frac{8}{N\pi^3}   \int_0^{\infty} d q \frac{g(q)^2 \cos^2\theta}{q} \int_0^{\infty} d y \frac{\sqrt{y}}{(y+g(q)^2)(y+\sin^2\theta)^2}\,.
\ee
Next using the Feynman parametrization trick we rewrite the integrand as:
\be
p(\theta) \approx \frac{8}{N\pi^3}  \int_0^{\infty} d q \frac{g(q)^2 \cos^2\theta}{q} \left(-\frac{\partial}{\partial (\sin^2 \theta)} \int_0^1 d t \int_0^\infty d y \frac{\sqrt{y}}{(g(q)^2 t + \sin^2\theta(1-t)+y)^2}\right)
\ee
With the help of a standard integral
\be
\int_0^{\infty} d y \frac{\sqrt{y}}{(a+y)^2} = \frac{\pi}{2\sqrt{a}}
\ee
we integrate over $y$ to get:
\be
p(\theta) \approx \frac{4}{N\pi^2} \int_0^{\infty} d q \frac{g(q)^2 \cos^2\theta}{q} \left(-\frac{\partial}{\partial (\sin^2\theta)} \int_0^1 d t \frac{1}{\sqrt{g(q)^2 t+\sin^2\theta(1-t)}}\right)\,.
\ee
Using
\be
\int_0^1 d t \frac{1}{\sqrt{a t + (1-t)}} = \frac{2}{1+\sqrt{a}}
\ee
one obtains:
\be
p(\theta) \approx \frac{4}{N\pi^2} \int_0^{\infty} d q \frac{g(q)^2 \cos^2\theta}{q} \left(-\frac{\partial}{\partial (\sin^2\theta)} \frac{2}{\sin\theta+g(q)}\right)\,,
\ee
which upon differentiation with respect to $\sin^2\theta$ gives:
\be
\label{eq:supp_gate_dis}
p(\theta) \approx \frac{4}{N\pi^2} \int_0^{\infty} d q \frac{g(q)^2 \cos^2\theta}{q \sin\theta (\sin\theta+g(q))^2}
\ee
To extract the leading behavior of $p(\theta)$ 
we split the integration into two regions: (i) $q < \tfrac{1}{2H}$ and (ii) $q > \tfrac{1}{2H}$. The two regions describe, respectively, the contributions of the lengthscales greater and smaller than the distance to the gate. In the region (i) we take a low-$q$ limit of $g(q)$ from Eq.\eqref{eq:gate_mod_simp} as:
\be
g(q) = g_0(1-e^{-2Hq}) \approx g_0 2 H q\,, \quad q \ll \frac{1}{2H}
\ee
Upon substitution into Eq.\eqref{eq:supp_gate_dis} this gives
\be
\label{eq:H_int_1}
\int_0^{\frac1{2H}} d q \frac{g_0^2 q(2H)^2}{\theta (\theta+g_0 2H q)^2} \approx \frac{1}{\theta}\ln\frac{g_0}{\theta}\,,
\ee
where we expanded $\cos\theta$ and $\sin\theta$ near $\theta\to0$ and focused on the leading order of divergence. For the region (ii) we take the approximation $g(q) \gg \sin(\theta)$ and thus:
\be
\label{eq:H_int_2}
\int_{\frac1{2H}}^{k_{\nu}} d q \frac{1}{q \theta} \approx \frac{1}{\theta}\ln(2H k_{\nu})
\ee
Combing both Eq. \eqref{eq:H_int_1} and Eq. \eqref{eq:H_int_2} we get
\be
p(\theta) \approx \frac{4}{N\pi^2} \frac{1}{\theta}\ln\frac{2H k_{\nu} g_0}{\theta}
\ee
exhibiting similar divergence to the angular distribution without the gate Eq.\eqref{eq:ang_dis_no_gate}, but with the $k_0$ cutoff replaced by $\frac1{2H}$.
We illustrate this by computing numerically an angular distribution for a system without a gate and the lower limit in the $q$ integral, Eq.\eqref{eq:final_ang}, replaced with $k_0^{\rm eff}=1/(2 H)$. The results are plotted in Fig. \ref{fig:app_N_vs_g}. For the plot in the main text we used $H = 10\,{\rm nm}$ (as $H = 300\,{\rm nm}$ was almost on top of the G/SiC curve) and we assumed as before the air/graphene/hBN/gate device parameters $\kappa_{\textrm{G/hBN}}=3.03$, $\kappa_{\textrm{air}}=1$, $\kappa_{\textrm{hBN}}=5.06$.

\end{document}